\documentclass[twocolumn,prl]{revtex4}
\usepackage{graphicx}
\usepackage{latexsym}

\begin{document}

\title{A Hybrid Al$_{0.10}$Ga$_{0.90}$As/AlAs Bilayer Electron System}
\author{E.\ P.\ De Poortere and M.\ Shayegan}
\address{Department of Electrical Engineering, Princeton University, Princeton, New Jersey 08544}
\date{\today}

\begin{abstract}
We have fabricated a device composed of two closely coupled
two-dimensional electron systems, one of which resides within an
AlAs quantum well (QW) at the $X$ point of the Brillouin zone
(BZ), while the other is contained at the $\Gamma$ point of the BZ
in the alloy Al$_{0.10}$Ga$_{0.90}$As, grown directly below the
AlAs QW. The electronic properties of these two systems are
strongly asymmetric: the respective cyclotron masses in the AlAs
and the Al$_{0.10}$Ga$_{0.90}$As layers, measured in units of the
free electron mass, are 0.46 and 0.07, while the effective
electron $g$-factors are approximately 8.5 and 0. With the help of
front and back gates, we can confine mobile carriers to either or
both of the two QWs, as confirmed by magnetotransport
measurements.
\end{abstract}

\pacs{}

\maketitle

In their approach to building a logic gate for quantum computing,
DiVincenzo et al. \cite{divincenzo99} have proposed a device in
which the quantum information unit (or qubit) is encoded in the
electron spin, which can be reversed by electrically shifting the
electron from a material with a low effective $g$-factor ($g^*$)
to a high-$g^*$ material in a magnetic field. The advantage of
this design lies in the fact that the spin coherence time is
typically longer than that of the spatial component of the
electron wavefunction; furthermore, the above design is in
principle compatible with existing semiconductor fabrication
techniques.

\begin{figure*}
\includegraphics[scale=.9]{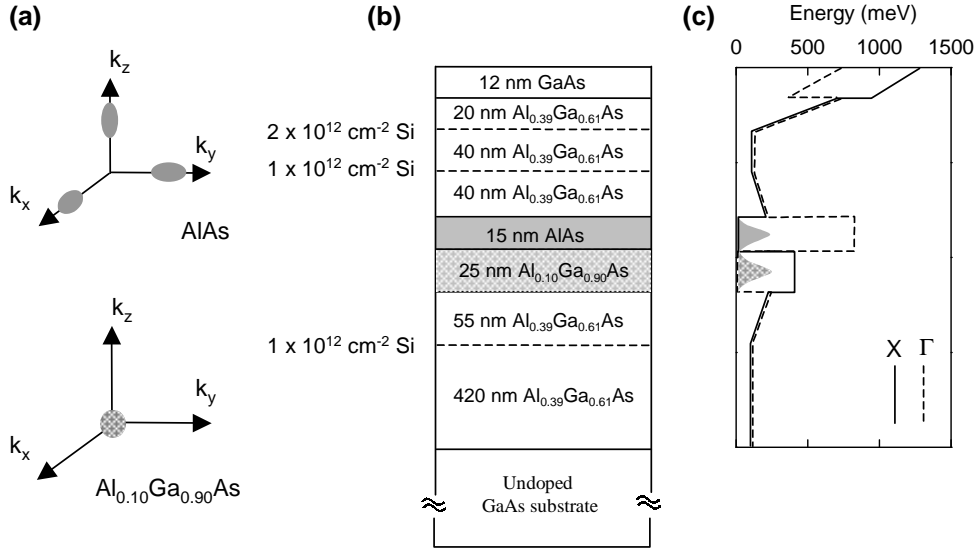}
\caption{(a) Fermi surfaces of conduction electrons in bulk AlAs
and Al$_{0.10}$Ga$_{0.90}$As. (b) Layer structure of the bilayer
sample. (c) Schematic conduction band diagram of the sample;
$X$-point and $\Gamma$-point conduction band edges are indicated
by solid and dashed lines, respectively. The sample is designed so
as to align the conduction band minima of the AlAs and AlGaAs QWs
(at the $X$ and $\Gamma$ points, respectively). Note that each QW
acts as a potential barrier for electrons in the other well.}
\label{structure}
\end{figure*}
Several groups have begun to implement DiVincenzo's scheme
\cite{salis01,jiang01,kato03}: in all of these cases, the electron
system is excited optically, and contained primarily in one
quantum well (QW). By contrast, the structure we present in this
work is composed of two separate QWs, made out of AlAs and
Al$_{0.10}$Ga$_{0.90}$As. Charge carriers in these QWs lie
respectively at the $X$ and $\Gamma$ points of the Brillouin zone,
at nearly identical conduction band energies ($E_c$). This
structure is therefore a direct implementation of DiVincenzo's
scheme \cite{divincenzo99}. We show through magnetoresistance
measurements that electrons in these two QWs possess different
$g$-factors: 8.5 and 0, respectively. Furthermore, the cyclotron
effective masses ($m^*$) of two-dimensional (2D) electrons in AlAs
and Al$_{0.10}$Ga$_{0.90}$As are 0.46 and 0.07, respectively
\cite{lay93,adachi85}. Our structure therefore allows to study the
properties of bilayer electron systems with unequal masses. We
also note that the $X$ and $\Gamma$ character of 2D electrons in
the two QW materials strongly suppresses tunnelling between the
two 2D electron layers \cite{tribe94}.

Our sample was grown by molecular beam epitaxy (MBE) on a GaAs
(100) substrate wafer. The structure of the epitaxial growth is
given in Fig.\ \ref{structure}(b), and contains a 15 nm-wide AlAs
QW grown atop a 25 nm-wide Al$_{0.10}$Ga$_{0.90}$As QW. Both QWs
are bordered by Al$_{0.39}$Ga$_{0.61}$As barriers, and the
Al$_{0.10}$Ga$_{0.90}$As layer was constructed as a $10 \times$
(1.85 nm GaAs/0.65 nm Al$_{0.39}$Ga$_{0.61}$As) superlattice. For
modulation doping, Si $\delta$-layers were inserted within the
barriers, separated from the QWs by 40 nm- and 55 nm-wide spacers
on the front and back sides, respectively. We designed the
structure so as to align the $X$ and $\Gamma$ $E_c$'s of the AlAs
and Al$_{0.10}$Ga$_{0.90}$As QWs, respectively
\cite{bandalignment}, as can be seen in the band diagram of Fig.\
\ref{structure}(c). We note that only the in-plane $X$ valleys of
the AlAs QW are occupied in our sample, in spite of their higher
out-of-plane effective mass, as the small stress resulting from
the AlAs-GaAs lattice mismatch raises the energy of the
out-of-plane valley \cite{lay93,vandestadt96}.

\begin{figure*}[t]
\includegraphics[scale=.8]{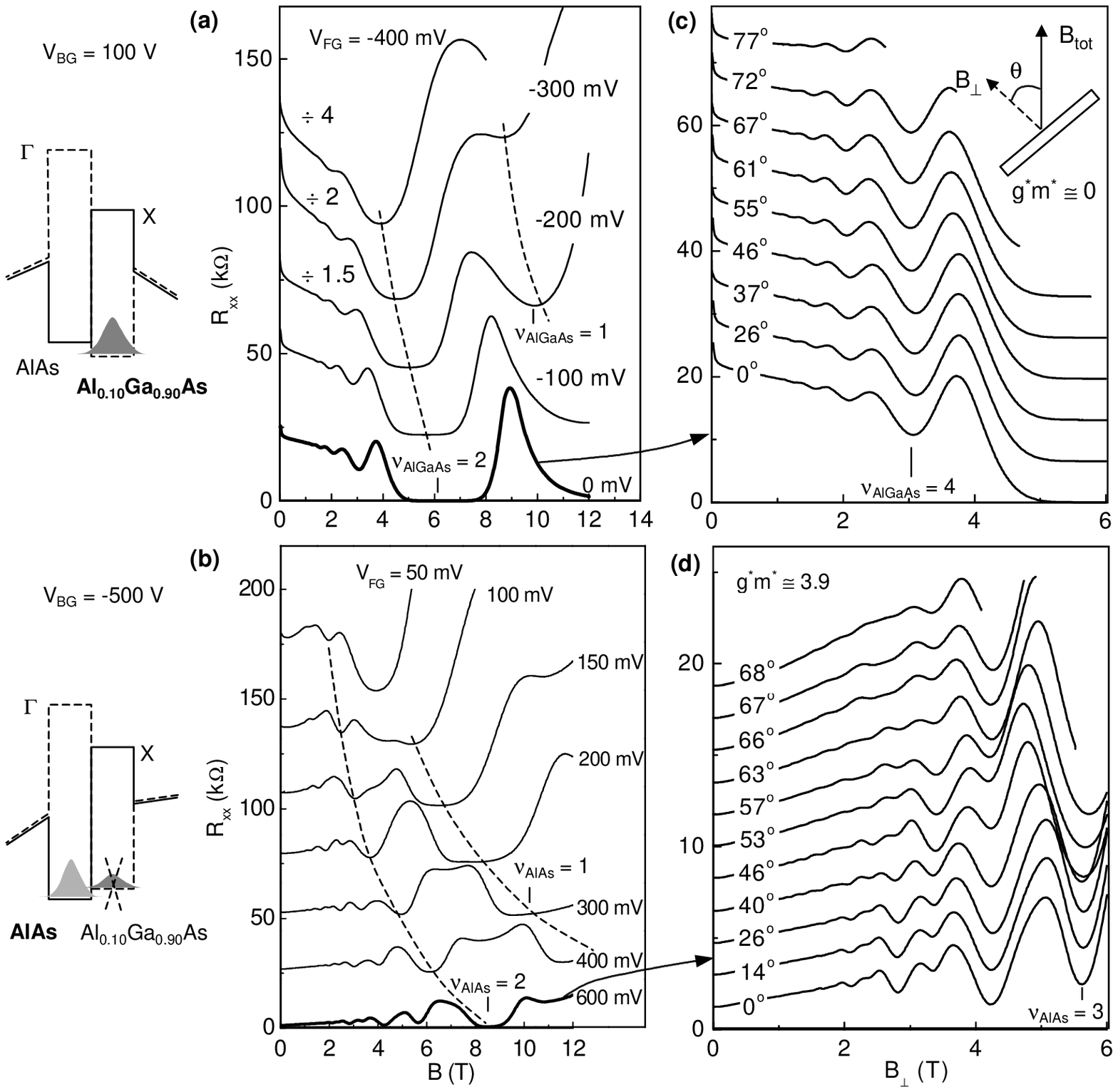}
\caption{Magnetoresistance of 2D electrons in the
AlAs/Al$_{0.10}$Ga$_{0.90}$As bilayer sample (traces are offset
vertically for clarity). In (a), $V_{BG} = 100$ V, and the front
gate bias ranges from -400 to 0 mV, values for which only the back
QW (Al$_{0.10}$Ga$_{0.90}$As) is occupied. Its density varies from
1.9 to 3.0$ \times 10^{11}$ cm$^{-2}$. In (b), both QWs are
filled, although most of the $R_{xx}$ minima arise from transport
through the AlAs layer. In this case, the 2D electron density in
the Al$_{0.10}$Ga$_{0.90}$As QW is $\sim$1.9$ \times 10^{11}$
cm$^{-2}$, while in the AlAs QW, the density ranges from 1.0 to
4.1$ \times 10^{11}$ cm$^{-2}$. (c) and (d): Tilted-field $R_{xx}$
data for two different gate biases, corresponding to the
highlighted traces in (a) and (b), respectively. $R_{xx}$ does not
vary with tilt angle in (c) ($g^* \simeq 0$), while in (d),
several coincidences occur ($g^* \simeq 8.5$). $g^*$ derived from
these data are consistent with electron transport occurring
through the Al$_{0.10}$Ga$_{0.90}$As layer in (c), and the AlAs
layer in (d).} \label{main}
\end{figure*}
For magnetotransport measurements, samples were patterned as Hall
bar mesas by optical lithography, and 140 nm-thick ohmic AuGeNi
contacts were deposited and alloyed at 460 $^o$C for 10 minutes in
forming gas. A semitransparent, 30 nm-thick, Ti/Au metal layer
deposited on the surface of each sample served as a front gate,
and an In-covered Si chip was soldered to the back of the sample
for back gate biasing. Transport measurements were performed in a
pumped $^3$He system at $T \sim 0.3$K and in magnetic fields up to
12 T.

In Fig.\ \ref{main}(a) and (b) we plot the longitudinal
magnetoresistance ($R_{xx}$) of the sample, measured as a function
of the perpendicular magnetic field, for different front- and
back-gate biases ($V_{FG}$ and $V_{BG}$, respectively). For the
traces shown in Fig.\ \ref{main}(a), $V_{FG}$ and $V_{BG}$ are
such that electrons populate the Al$_{0.10}$Ga$_{0.90}$As QW only
(``AlGaAs regime''). Aside from an $R_{xx}$ minimum at Landau
level filling $\nu_{AlGaAs} = 1$ \cite{nudef}, $R_{xx}$ minima are
strong for even filling factors only. This behavior is expected
for Al$_{0.10}$Ga$_{0.90}$As 2D electrons, for which $g^* \sim 0$
\cite{shukla00}. In Fig.\ \ref{main}(b), the 2D electron density
in the Al$_{0.10}$Ga$_{0.90}$As QW has been reduced to 1.9 $\times
10^{11}$ cm$^{-2}$ by application of a large negative $V_{BG} =
-500$ V. In this state, increasing $V_{FG}$ acts to raise the
carrier density in the AlAs QW (``AlAs regime''). The minima
observed in the $R_{xx}$ data all correspond to integer Landau
levels fillings of the AlAs 2D electrons (both even and odd
$\nu$'s are apparent). Furthermore, the data in this regime look
highly similar to those measured in single AlAs QWs of comparable
quality \cite{lay93}. $R_{xx}$ data in Fig.\ \ref{main}(b) also
show that for $V_{FG}
> 150$ mV, a large fraction of the 10 nA current driven through the sample passes through the AlAs QW, as none of
the $R_{xx}$ oscillations in Fig. \ref{main}(b) can be associated
with the AlGaAs 2D electrons.

As further evidence that both the Al$_{0.10}$Ga$_{0.90}$As and
AlAs QWs can be occupied with charge carriers, we measure the
effective $g^*$ of the bilayer system for two different sets of
front- and back-gate biases, corresponding to the highlighted data
in Fig.\ \ref{main}(a) and (b). We derive $g^*$ using the
coincidence method \cite{fang68}: in this technique, the sample is
placed at an angle $\theta$ with respect to the direction of the
$B$ field (Fig.\ \ref{main}(c), inset). For particular $\theta$
values, the Zeeman energy, $E_Z = g^*\mu_B B$ (where $\mu_B$ is
the Bohr magneton), is an integer multiple of the cyclotron
energy, equal to $\hbar eB_{\bot}/m^*m_e$ ($B_{\bot}$ is the
perpendicular component of $B$ and $m_e$ is the bare electron
mass): these two energies depend on the total and perpendicular
magnetic field, respectively, whose ratio is controlled by cos
$\theta$. In these conditions, the energy ladders of spin-up and
spin-down Landau levels overlap, resulting in the weakening of
even (or odd) $R_{xx}$ quantum Hall minima. In our case, the value
of $\theta$ for which the $\nu = 3$, 5, 7,... $R_{xx}$ minima
disappear or become weakest is related to $g^*$ and to $m^*$ via
the formula $g^*m^* = 4$cos $\theta$ \cite{papadakis99}.

We concentrate on the two highlighted traces in Fig.\
\ref{main}(a) and (b). For each of these states, we then measure
$R_{xx}(B)$ for different $\theta$, as plotted in Fig.\
\ref{main}(c) and (d). In the first case, the $R_{xx}$ data is
insensitive to $\theta$, showing that $g^*m^* \simeq 0$. In the
latter case [Fig.\ \ref{main}(d)], Landau level coincidences occur
for $\theta$ = 14$^o$, 53$^o$, and $63^o$, yielding $g^*m^* =
3.9$. The measured $g^*$ are therefore $\simeq$ 0 in the former
case, and 8.5 in the latter (taking $m^* = 0.46$ \cite{lay93}).
$g^* \simeq 0 $ is the value expected for Al$_x$Ga$_{1-x}$As 2D
electrons with $x \sim 0.10$ \cite{shukla00}, while $g^* = 8.5$ is
close to the value previously obtained for AlAs 2D electrons
\cite{papadakis99}. These measurements confirm that the $R_{xx}$
data in the upper panels of Fig.\ \ref{main} describe
Al$_{0.10}$Ga$_{0.90}$As 2D electrons, while the lower panels of
the same figure correspond to AlAs 2D electrons.

\begin{figure*}
\includegraphics[scale=.35]{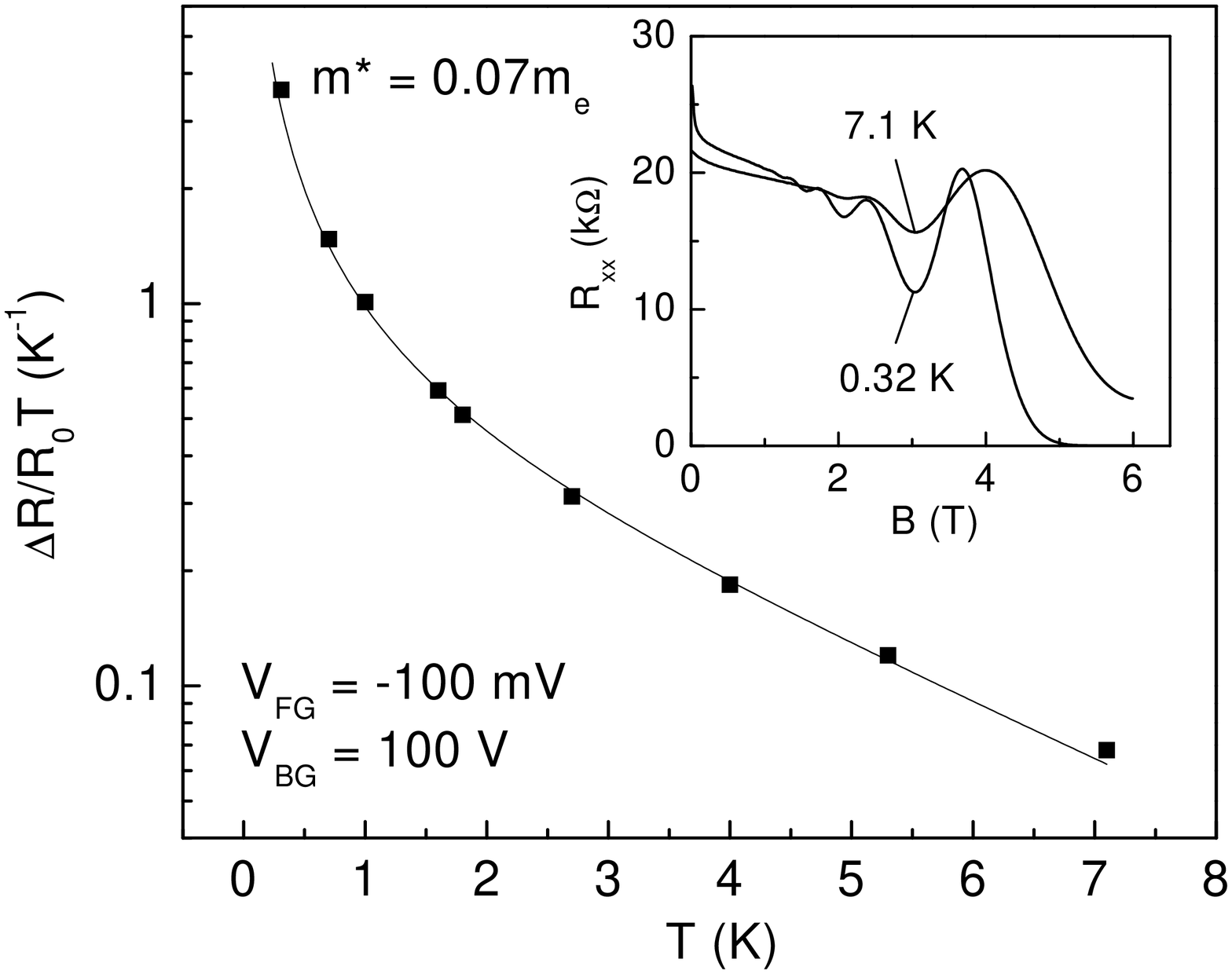}
\caption{Relative amplitude of the Shubnikov-de Haas
magnetoresistance oscillations at $B = 3$ T plotted vs.\ $T$, and
$R_{xx}$ vs.\ $B$ data (inset), for the state ($V_{FG}$ = -100 mV)
of Fig.\ \ref{main}(a). A fit to the Dingle formula yields $m^* =
0.07$, consistent with the effective mass of
Al$_{0.10}$Ga$_{0.90}$As 2D electrons.} \label{AlGaAsmass}
\end{figure*}

Additionally, the Dingle plot of Fig.\ \ref{AlGaAsmass}, obtained
from the temperature dependence of the Shubnikov-de Haas
oscillations of the $V_{FG} = -100$ mV data in Fig.\
\ref{main}(a), yields $m^* = 0.07$, the effective mass expected
for Al$_{0.10}$Ga$_{0.90}$As electrons. Our Dingle data in the
AlAs regime are so far harder to interpret, but yield a range of
effective masses all somewhat larger than $\simeq 0.5$, also
consistent with the 2D electron mass in AlAs.

We conclude with practical considerations on sample fabrication.
Samples similar to the one we have described require a precise
alignment between the electron energy levels in AlAs and
Al$_{0.10}$Ga$_{0.90}$As, and thus demand: (1) knowledge of the
conduction band offsets between the two materials and (2) good
control over Al and Ga fluxes during epitaxial growth. Such
control is difficult but possible to achieve in practice. In our
case, since we do not rotate the wafer during MBE growth, we have
a flux inhomogeneity of about $\pm 15$ \% across the two-inch
wafer. This spatial variation results in a slow variation of the
QW thickness and of the Al concentration gradient within the
Al$_{0.10}$Ga$_{0.90}$As QW across the wafer, and increases the
likelihood that in some section of the wafer both QWs will be
occupied. With this technique, we have so far obtained four other
samples (from three different wafers) similar to the one presented
in this article. We add that present day MBE technology makes it
possible to grow very uniform and homogeneous wafers with precise
composition and thicknesses. Using our sample parameters (Fig.\ 1)
as a starting point, and by rotating the wafer during MBE growth,
it should be possible to grow wafers that contain larger areas of
useful samples.

In summary, we have fabricated a new type of modulation-doped
heterostructure, in which two 2D electron systems reside
side-by-side in two different materials, AlAs and
Al$_{0.10}$Ga$_{0.90}$As, with widely different effective
$g$-factors and masses \cite{vakili03}. This structure allows to
modulate electrostatically the $g$-factor of 2D carriers from a
vanishingly small to a large value, and offers the first step
towards the realization of a spin-based quantum computer. In
addition, it constitutes a bilayer 2D system with near-zero
tunnelling, but small separation between the component 2D electron
gases, a system in which new electronic ground states may be
observed. A superconducting transition has for example been
predicted in such a two-mass bilayer 2D system \cite{macdonald03}.

This work is supported by the NSF and ARO.


\begin{references}

\bibitem{divincenzo99}
D.\ P.\ DiVincenzo, G.\ Burkard, D.\ Loss, and E.\ V.\ Sukhorukov,
http://xxx.lanl.gov/cond-mat/9911245 (1999).

\bibitem{salis01}
G.\ Salis {\it et al.}, Nature {\bf 414}, 619 (2001). For
measurements of the 2D electron effective mass in parabolic QWs,
see also X.\ Ying, K.\ Karra\"{i}, H.\ D.\ Drew, M.\ Santos, and
M.\ Shayegan, Phys.\ Rev.\ B {\bf 46}, 1823 (1992).

\bibitem{jiang01}
H.\ W.\ Jiang and E.\ Yablonovitch, Phys.\ Rev.\ B {\bf 64},
041307 (2001).

\bibitem{kato03}
Y.\ Kato {\it et al.}, Science {\bf 299}, 1201 (2003).

\bibitem{lay93}
T.\ S.\ Lay {\it et al.}, Appl.\ Phys.\ Lett. {\bf 62}, 3121
(1993).

\bibitem{adachi85}
S.\ Adachi, J.\ Appl.\ Phys.\ {\bf 58}, R1 (1985).

\bibitem{tribe94}
W.\ R.\ Tribe, P.\ C.\ Klipstein, and G.\ W.\ Smith, in {\it
Proceedings of the 22nd Int.\ Conf.\ on the Physics of
Semiconductors}, edited by D.\ J.\ Lockwood (World Scientific,
Singapore, 1995), p.\ 759.

\bibitem{bandalignment}
More precisely, we attempted to align the {\it first energy
levels} of the conduction bands within each QW, which, because of
QW confinement, are slightly higher than the corresponding
$E_c$'s.

\bibitem{vandestadt96}
A.\ F.\ W.\ van de Stadt, P.\ M.\ Koenraad, J.\ A.\ A.\ J.\
Perenboom, and J.\ H.\ Wolter, Surf.\ Sci.\ {\bf 361-362}, 521
(1996) and references therein.


\bibitem{nudef}
The Landau level filling (or ``filling factor'') is defined as the
number of quantized Landau levels occupied by the 2D electrons in
a magnetic field.

\bibitem{shukla00}
S.\ P.\ Shukla {\it et al.}, Phys.\ Rev.\ B {\bf 61}, 4469 (2000).


\bibitem{fang68}
F.\ F.\ Fang and P.\ J.\ Stiles, Phys.\ Rev.\ {\bf 174}, 823
(1968).

\bibitem{papadakis99}
S.\ J.\ Papadakis, E.\ P.\ De Poortere, and M.\ Shayegan, Phys.\
Rev.\ B {\bf 59}, R12743 (1999).

\bibitem{vakili03}
Recently also, a bilayer AlAs QW structure with asymmetric
electron masses and $g$-factors has been demonstrated by K.\
Vakili, Y.\ P.\ Shkolnikov, E.\ Tutuc, E.\ P.\ De Poortere, and
M.\ Shayegan [http://xxx.lanl.gov/cond-mat/0309385 (2003)]. In
this device, electrons in one AlAs QW populate the {\it in-plane}
($X_{xy}$) valley, while electrons in the other QW reside in an
{\it out-of-plane} ($X_z$) valley.

\bibitem{macdonald03}
E.\ Bascones, T.\ Jungwirth and A.H.\ MacDonald, unpublished.




\end{references}
\end{document}